\documentstyle[psfig,nicV]{article}

\def\cs{cm$^{-2}$ s$^{-1}$}
\def\ga{$\gamma$}
\def\aa{$\alpha$}
\def\coa{$^{56}$Co}
\def\cob{$^{57}$Co}
\def\nia{$^{56}$Ni}
\def\fea{$^{56}$Fe}
\def\feb{$^{57}$Fe}
\def\fec{$^{60}$Fe}
\def\nib{$^{57}$Ni}
\def\alb{$^{26}$Al}
\def\tii{$^{44}$Ti}
\def\mg{$^{26}$Mg}
 \def\nh{$^{60}$Ni}
\def\na{$^{22}$Na}
\def\co{$^{56}$Co}
\def\ci{$^{57}$Co}
\def\ch{$^{60}$Co}
\def\fe{$^{56}$Fe}
\def\fr{$^{57}$Fe}
\def\fh{$^{60}$Fe}
\def\ne{$^{22}$Ne}
\def\ti{$^{44}$Ti}
\def\sca{$^{44}$Sc}
\def\ca{$^{44}$Ca}

\def\ra{$\rightarrow$}
\def\aa{$\alpha$}

\newcommand{\chem}[2]{$\rm{}^{#1}\kern-0.8pt#2$}
\newcommand{\chim}[2]{\rm{}^{#1}\kern-0.8pt#2}
\newcommand{\reac}[6]{$\rm\,{}^{#1}\kern-0.8pt{#2}\,({#3}\,,{#4})\,
           {}^{#5}\kern-0.8pt{#6}\,$}
\newcommand{\gsimeq}{\,\,\raise0.14em\hbox{$>$}\kern-0.76em\lower0.28em\hbox  
{$\sim$}\,\,}
\newcommand{\lsimeq}{\,\,\raise0.14em\hbox{$<$}\kern-0.76em\lower0.28em\hbox  
{$\sim$}\,\,}

\newcommand{\ms}{$\rm M_{\odot}$}
\newcommand{\zs}{$\rm Z_{\odot}$}
\newcounter{obj}

\begin{document}

\heading{%
Cosmic radioactivities
}

\par\medskip\noindent

\author{Marcel Arnould$^1$ and Nikos Prantzos$^2$}

\address{Institut d'Astronomie et d'Astrophysique, Universit\'e
Libre de Bruxelles \\
C.P. 226, Bd. du Triomphe, B-1050 Brussels, Belgium}
\address{Institut d'Astrophysique de Paris, 98bis Bd. Arago
F-75014 Paris, France}

\begin{abstract}
Radionuclides with half-lives ranging from some years to billions of years
presumably synthesized outside of the solar system are now recorded in ``live" or
``fossil" form in various types of materials, like meteorites or the galactic cosmic
rays. They bring specific astrophysical messages the deciphering of which is briefly
reviewed here, with special emphasis on the contribution of Dave Schramm and his
collaborators to this exciting field of research. Short-lived radionuclides
are also present in the Universe today, as directly  testified by the $\gamma$-ray
lines emitted by the de-excitation of their daughter products. A short review of
recent developments in this field is also presented.

\end{abstract}

\section{Introduction}
When Marie and Pierre Curie discovered just one century ago the mysterious
phenomenon of radioactivity, they certainly did not imagine that they had 
opened the way to a better understanding of many facets of the Universe! Dave
Schramm has entered this way, clearing many of its misty sections with his uncommon
enthusiasm and energy, and with a remarkably vivid view of the astrophysical
potentialities along the road. This contribution is a tribute to this aspect of 
Dave's mutiple activities, and in particular to his important contribution to the
unravelling of the astrophysical messages from long- and short-lived
radionuclides. We will in particular briefly review some selected aspects of
nucleo-cosmochronology, as well as of the field of the isotopic anomalies of
radionuclidic origin. Additionally, the decay of certain radionuclides manifests
itself quite spectacularly through the emission of
de-excitation lines in the $\gamma$-ray domain. The study of these radionuclides
offers invaluable information on their production sites and is the subject of
$\gamma$-ray line astronomy, a recently developed astrophysical discipline
which we review in the last section of this paper.

In contrast, we will not deal here with the very interesting subject of the
production of a large variety of radionuclides in terrestrial or solar-system
extra-terrestrial matter bombarded by galactic cosmic rays or solar
energetic particles. This spallative production is of importance for the deciphering
of the record of these energetic particles, and in particular of the time variations
of their fluxes, but its interest goes well beyond astrophysics or
planetology. The reader is referred to e.g. \cite{Michel98} for a review and
some references.

\section{Cosmochronometry}
 
The dating of the Universe and of its various constituents, referred to as
``cosmochronology'', is one of the tantalizing tasks in modern science. 
This field is in fact concerned with different ages, each  one of them corresponding to
an epoch-making event in the past (e.g.~\cite{Flam90} for many contributions on this
subject, and especially Dave's contribution \cite{Schramm90}). They are in
particular the age of the Universe $T_{\rm U}$, of the  globular clusters $T_{\rm
GC}$, of the Galaxy [as (a typical?) one of many  galaxies] $T_{\rm G}$, of the
galactic disc
$T_{\rm disc}$, and of the  non-primordial nuclides in the disc $T_{\rm nuc}$, with
$T_{\rm U}
\gsimeq  T_{\rm GC} \approx$ ($\gsimeq$?) $T_{\rm G} \gsimeq T_{\rm disc} \approx
T_{\rm nuc}$. As a consequence, cosmochronology involves not only
cosmological models and observations, but also various other astronomical and
astrophysical studies, and even invokes some nuclear physics information. 

The cosmological models can help determining  $T_{\rm U}$, as well as, to some
extent at least, 
$T_{\rm GC}$ and $T_{\rm disc}$ 
(e.g.~\cite{Arnould90,Fowler86,Tayler86} for  brief 
accounts). The $T_{\rm GC}$ or $T_{\rm disc}$ values have also been evaluated from
the use of the Herztsprung-Russell diagram (HRD) (e.g.~\cite{Jimenez98a,VandenBerg97}
for recent reviews), or of so-called ``luminosity functions,'' which provide the
total number of stars per absolute magnitude interval as a function of absolute
magnitudes. In particular, the luminosity function of white-dwarf stars has been
proposed as a priviledged $T_{\rm disc}$ evaluator (e.g.
\cite{Hernanz94,Oswalt96}). Nucleo-cosmochronological techniques have also been
developed in order to evaluate
$T_{\rm nuc}$, and are briefly discussed below. 
Each of these methods has advantages and weaknesses of its own, as briefly reviewed
by e.g.~\cite{Arnould90}. 
%
 
\subsection{``Long-lived" nucleo-cosmochronometers: generalities} 

The dating method that most directly relates to nuclear astrophysics is 
referred to as ``nucleo-cosmochronology.'' It primarily aims at determining
the age $T_{\rm nuc}$ of the nuclides in the galactic disc through the use of
the observed bulk (meteoritic)  abundances of radionuclides with lifetimes
commensurable with presumed $T_{\rm disc}$ values (referred to in the following as
``long-lived" radionuclides). Consequently, it is hoped to provide at least a lower
limit to $T_{\rm disc}$. The most studied chronometries involve \chem{187}{Re} or the
trans-actinides \chem{232}{Th}, \chem{235}{U} and \chem{238}{U}.    

In order to establish a good chronometry based on these radioactive nuclides, one
needs to have firstly a good set of input data concerning (isotopic) abundances
and nucleosynthesis yields, in addition to the radioactive half-lives. Another issue
concerns the necessity, and then the possibility, of using detailed models for the
chemical evolution of the Galaxy in order to gain a reliable
nucleo-cosmochronological information if indeed the bulk solar-system composition
witnesses the perfect mixing of a large number of nucleosynthetic events. The status
of these various requirements is briefly examined in the following sections for
several cosmic clocks.
\subsubsection{The trans-actinide clocks} 

The familiar long-lived   
\chem{232}{Th}-\chem{238}{U} and \chem{235}{U}-\chem{238}{U} chronometric pairs
\cite{Fowler60} are developed on grounds of their abundances at the time of
solidification in the solar system some $4.56 \times 10^9$ y ago. This information
is obtained by extrapolating back in time the present meteoritic content of these
nuclides. If the so-derived abundances are affected by some uncertainties, these
are not, however, the main problems raised when attempting to use these
radionuclides as reliable nuclear clocks.  Their usefulness in this respect indeed
depends in particular on the availability of precise production ratios. Such
predictions at the level of accuracy needed for getting a truly useful chronometric
information are out of reach at the present time. One is indeed dealing with nuclides
that can be produced by the r-process only, which suffers from very many
astrophysics and nuclear physics problems, in spite of much effort by many
researchers, including Dave and his collaborators (e.g.
\cite{Meyer94} for relevant references). The r-process problems are particularly acute
for the Th and
U isotopes referred to above. They are indeed the only naturally-occuring
nuclides beyond \chem{209}{Bi}, so that any extrapolation relying on semi-empirical
analyses and fits of the solar r-process abundance curve is in danger of being
especially unreliable. The difficulty is further reinforced by the fact that
most of the r-process precursors of U and Th are nuclei that are unknown in
the laboratory, and will remain so for a long time to come.  Theoretical
predictions of properties of relevance, like masses, 
$\beta$-decay strength functions and fission barriers, are extremely
difficult, particularly as essentially no calibrating points exist.
This problem would linger even if a realistic r-process model were given, which is
not the case at the present time (e.g. \cite{Arnouldiop}). Last but not least, most
of the tremendous amount of work devoted in  the past to the trans-actinide
chronometry   has adopted simple functionals for
the time dependence of the r-process nucleosynthesis rate  (a.k.a. ``Mickey Mouse
Models'' coined by Pagel \cite{Pagel90}) with little consideration of the chemical
evolution in the solar neighbourhood. This view, which originated almost 4 decades
ago \cite{Fowler60}, has had (and still has) a few sympathisers indeed (e.g.
\cite{Cowan91}; also \cite{Arnouldiop} for some references). 

The necessity of the development of the long-lived chronometers in the framework of
models for the chemical evolution of the Galaxy has been first pointed out by
Tinsley \cite{Tinsley77}. The introduction of nucleo-cosmochronological
considerations in such models is not a trivial matter, however. The intricacies come
in particular from ``astration" effects, which have to do with the fate of the
chronometers once absorbed from the interstellar medium by the stars at their birth
(e.g. \cite{Yokoi83}). However, Dave, in collaboration with Wasserburg
\cite{Schramm70}, has made an important contribution to nucleo-cosmochronology by
showing that one can make the economy of these chemical evolution models as
long as a mere determination of {\it age limits} could satisfy one's curiosity. This
interesting so-called ``model-independent approach" has led to the conclusion
that $9 \lsimeq T_{\rm nuc} \lsimeq 27$ Gyr \cite{Meyer86}.
 
There has also been an attempt to develop a Th-chronometry \cite{Pagel89}
on grounds of the relative abundances of Th and Eu (which is
presumed to be dominantly produced by the r-process) observed at the surface of
stars with various metallicities.\footnote{Originally, an attempt was made to use
the observed Th/Nd ratios \cite{Butcher87}, albeit the disadvantage of Nd being
possibly produced also by the s-process}
Under the assumption, which may sound reasonable but has not at all to be
taken for granted, that any r-process in the past has produced Th and Eu
with a constant solar-system ratio, the age determination is reduced to the problem
of mapping the metallicity on time through a chemical evolution model.
High-quality observational Th/Eu abundance data in stars of various
metallicities are accumulating \cite{daSilva90,Francois93,Sneden96}. In
spite of some attempts \cite{Cowan97}, much remains to be done in the difficult task
of deriving $T_{\rm nuc}$ from these observations.

The Th-chronometry could be put on safer grounds if the Th/U ratios  
would be known in a variety of stars with a high enough accuracy. These nuclides
are indeed likely to be produced simultaneously, so that one may hope to be able to
predict their production ratio more accurately than the Th/Eu one. Even in such
relatively favourable circumstances, one would still face the severe question of
whether Th and U were produced  in exactly the same ratio in presumably a few
r-process events (a single one?) that have contaminated the material from
which metal-poor stars formed. Even if this ratio turns out to be the same
indeed, its precise value remains to be calculated (see e.g.
\cite{Arnould90a} for an illustration of the dramatic impact of a variation
in the predicted Th/U ratio on predicted ages).
\subsubsection{The \chem{187}{Re} - \chem{187}{Os} chronometry} 

First introduced by Clayton \cite{Clayton64}, the chronometry using the
\chem{187}{Re} - \chem{187}{Os} pair is able to avoid the difficulties 
related to the r-process modelling. True, \chem{187}{Re} is an r-nuclide.
However, \chem{187}{Os} is not produced  directly by the r-process, but 
indirectly via the $\beta^-$-decay of \chem{187}{Re} ($t_{1/2} 
\approx 43$ Gy) over the galactic lifetime. This makes it in principle
possible to derive a lower bound for $T_{\rm nuc}$
from the mother-daughter abundance ratio, provided that the ``cosmogenic"
\chem{187}{Os} component is deduced from the solar abundance by subtracting
its s-process contribution. This chronometry is thus in the first instance
reduced to a question concerning the s-process. Other good news come from the
recent progress made in the measurement of the abundances of the concerned
nuclides in meteorites (e.g.~\cite{Faestermann98} for references). This input
is indeed essential also for the establishment of a reliable chronometry.
 
Although the s-process is better understood than the r-process, this
chronometry is facing specific problems. They may be summarized as follows
(see e.g.~\cite{Takahashi98a} for a short account): 1) the evaluation of the
\chem{187}{Os} s-process component from the ratio of its production to the one
of the s-only nuclide \chem{186}{Os} is not a trivial matter, even in the
simple local  steady-flow approximation (constancy of the product of the
abundances by the stellar neutron capture rates over a restricted
$A$-range). The difficulty relates to the fact that the \chem{187}{Os} 9.75
keV excited state can contribute significantly to the stellar neutron capture
rate because of its thermal population in s-process conditions ($T \gsimeq
10^8$ K) (e.g.~\cite{Winters86,Woosley79}). The ground-state capture rate
measured in the laboratory has thus to be modified by a theoretical
correction. In addition, the possible branchings of the s-process path in
the $184 \leq A \leq 188$ region may be responsible of a departure from the
steady-flow predictions for the \chem{187}{Os}/\chem{186}{Os} production ratio
(e.g.~\cite{Arnould84,Kaeppeler91}); and 2) at the high temperatures, and thus
high ionisation states, \chem{187}{Re} may experience in stellar interiors,
its $\beta$-decay rate may be considerably, and sometimes enormously, enhanced
over the laboratory value by the bound-state $\beta$-decay of its ground
state to the 9.75 keV excited state of \chem{187}{Os} (e.g.~\cite{Yokoi83}).
Such an enhancement has recently been beautifully confirmed by the
measurement of the decay of fully-ionised \chem{187}{Re} at the GSI storage
ring \cite{Bosch96,Kienle98}. The inverse transformation of \chem{187}{Os}
via free-electron captures is certainly responsible for additional
corrections to the stellar \chem{187}{Re}/\chem{187}{Os} abundance
ratio (e.g.~\cite{Arnould72,Yokoi83}). Further complications arise
because these two nuclides can be concomitantly destroyed by
neutron captures in certain stellar locations \cite{Yokoi83}.
 
All the above effects have been studied in the framework of realistic  
evolution models for $1 \lsimeq M \lsimeq 50$ M$_\odot$ stars and of a galactic
chemical evolution model that is constrained by observational data in the solar
neighbourhood \cite{Takahashi98a,Takahashi98}. This work, which is an up-date of
\cite{Yokoi83} with regards to meteoritic abundances, nuclear input data, stellar
evolution models and observational constraints, concludes that $T_{\rm nuc} \approx 15
\pm 3$ Gy. Even lower ages of about 9 Gy, as derived from the model-independent
approach \cite{Schramm70,Schramm90} (Sect.~2.1.1), can not conclusively be excluded
within the remaining uncertainties in the chemical evolution model parameters.

These results may imply that the \chem{187}{Re} - \chem{187}{Os} chronometry has not
yet much helped narrowing the age range derived from other methods. There is still
ample room for improvements, however, and there is reasonable hope that the Re - Os
chronometry will be able to set some meaningful limits on $T_{\rm nuc}$ in a near
future, and independently of other methods.

\subsubsection{\chem{\rm 176}{\rm Lu}, a long-lived s-process radionuclide}

The long-lived \chem{176}{Lu} ($t_{1/2} = 41$ Gy) has the remarkable property of
being shielded from the r-process, and thus to be a pure s-process product. It has
been proposed by Dave and his collaborators \cite{Audouze72}, and independently by
\cite{Arnould73}, to be a potential chronometer for the s-process, the other
long-lived radionuclides probing the r-process instead. These early works pointed out
some possible uncertainties in the solar \chem{176}{Lu} abundance, as well as in its
production predicted from s-process models. The latter problem relates directly to
the branching in the s-process path due to the 125 keV \chem{176}{Lu^{\rm m}}
isomeric state. More specifically, the two different paths
\chem{175}{Lu}(n,$\gamma$)\chem{176}{Lu^{\rm
g}}(n,$\gamma$)\chem{177}{Lu}($\beta^-$)\chem{177}{Hf} and
\chem{175}{Lu}(n,$\gamma$)\chem{176}{Lu^{\rm
m}}($\beta^-$)\chem{176}{Hf}(n,$\gamma$)\chem{177}{Hf} may well develop during a
s-process (\chem{176}{Lu^{\rm g}} designates the \chem{176}{Lu} ground state). The
resulting
\chem{176}{Lu^{\rm g}}/\chem{176}{Hf} production ratios depend on the relative
importance of these two branchings, and thus mainly on the relative population of
the ground and isomeric \chem{176}{Lu} states. Two limiting situations are relatively
simple to handle. The first one is obtained if \chem{176}{Lu^{\rm g}} and
\chem{176}{Lu^{\rm m}} have no time in a given astrophysical environment for being
connected electromagnetically. This situation is made plausible by the 
large difference in the spin and $K$ quantum
number of the two states. In such conditions, the relative importance of the two
s-process branches is just given by the \reac{175}{Lu}{n}{\gamma}{176}{Lu^{\rm
m}}/\reac{175}{Lu}{n}{\gamma}{176}{Lu^{\rm g}} cross section ratio, the value of
which can be obtained from experiments. The other extreme is obtained if 
\chem{176}{Lu^{\rm g}} and \chem{176}{Lu^{\rm m}} are coupled electromagnetically
strongly enough for the relative populations of these two states to be
``thermalized", i.e. follow the rules of statistical equilibrium. In such
conditions, the relative importance of the two s-process branches is essentially
governed by temperature, as is the effective decay rate of the thermalized
\chem{176}{Lu}. 

Since the pioneering studies mentioned above, much work has been devoted to the
question of the possibility of thermalization of the \chem{176}{Lu} isomeric and
ground states in astrophysical plasmas, and to the measurement of the neutron
capture cross sections needed for the calculation of the s-process
\chem{176}{Lu}/\chem{176}{Hf} production ratio (e.g. \cite{Klay91,Lesko91}, and
references therein). From these efforts, it is generally
concluded to-day that the \chem{176}{Lu^{\rm g}} s-process yields are so sensitive
to temperatures and neutron densities that they cannot be evaluated precisely enough
for chronological purposes. Instead, \chem{176}{Lu^{\rm g}} could rather be
considered as a s-process thermometer.
\section{The message from extinct ``short-lived" radionuclides}

The discovery of isotopic anomalies attributed to the decay in some
meteoritic material of now extinct radionuclides with half-lives in the approximate
$10^5 \lsimeq t_{1/2} \lsimeq 10^8$ y range (referred to in the following as
``short-lived" radionuclides) has broadened the original astrophysical interest for
cosmic radioactivities. Even the ``ultra-short" radionuclides \chem{22}{Na} ($t_{1/2}
= 2.6$ y) and \chem{44}{Ti} ($t_{1/2} \approx 60$ y,
e.g. \cite{Wietfield99}) are likely to have left
their signatures in some meteorites. The interpretation of the message from these
anomalies has been the focus of much work and excitement.

One important issue raised by the extinct radionuclides concerns their
presence in the early solar system in ``live" form, or just in the form of their
daughter products (``fossils"). In the first case, the anomalies have of course to be
located in solar-system indigenous solids, while they have to be found in alien
(presolar) material in the second situation. At present, there is clear evidence that
meteorites contain both live and fossil signatures of short-lived nuclides, and
the messages they carry are quite different indeed. In contrast, the meteoritic
content of the ultra-short-lived nuclides has obviously to be of fossil nature, in
view of the lifetimes involved. 

\subsection{Live short-lived radionuclides in the early solar system}
 
At the end of the sixties, Dave and his collaborators \cite{STW70} have
contributed in an important way to the pioneering searches for the signatures of
extinct radionuclides in meteorites (e.g. \cite{Wasserburg82} for a historical
account) by establishing techniques for the high precision measurement of the Mg
isotopic composition in order to search for \chem{26}{Mg} excesses due to the
\chem{26}{Al} decay in meteorites of different types and in lunar samples. From this
study, it was concluded that the upper limits on the
(\chem{26}{Al}/\chem{27}{Al})$_0$ ratio\footnote{Here and in the following, the
subscript 0 refers to the start of solidification in the solar system some $4.56
\times 10^9$ y ago} in the analyzed materials was ranging from well below $10^{-6}$
to about $2 \times 10^{-6}$. It is established by now that \chem{26}{Al} has been
live in the solar system at a canonical level of (\chem{26}{Al}/\chem{27}{Al})$_0
\approx 2 \times 10^{-5}$
\cite{MacPherson95}. Persuasive evidence for the existence of other live
radionuclides has accumulated, and concerns nowadays \chem{53}{Mn}, \chem{60}{Fe},
\chem{107}{Pd}, \chem{129}{I}, \chem{146}{Sm} and \chem{244}{Pu}. This is also likely
the case for \chem{182}{Hf} and \chem{41}{Ca}, the presence of which has recently
been found to be correlated with the one of \chem{26}{Al} in some primitive
meteorites \cite{Sahijpal98}. Some weaker evidence has been gathered
about \chem{36}{Cl}, \chem{92}{Nb}, \chem{99}{Tc} and \chem{205}{Pb} (see e.g.
\cite{Podosek97} for a review and references).
 
The demonstrated existence of short-lived radionuclides in live form in the early
solar system can usefully constrain the chronology of the nebular and planetary
events at that epoch (e.g. \cite{Podosek97} for details). From a more astrophysical
point of view, these observations are generally considered to provide the most
sensitive radiometric probes concerning discrete nucleosynthesis  events that
presumably contaminated the solar system at times between about $10^5$ and
$10^8$ y prior to the isolation of the solar material from the general galactic
material. Of course, this statement assumes implicitly that the radionuclides of
interest have not been synthesized in the solar system itself. Can such a local
production scenario be rejected right the way ? Clearly, a large variety of
radionuclides are produced continuously in the present extraterrestrial solar
system matter, as well as in terrestrial samples, by spallation reactions induced by
Galactic Cosmic Rays (GCRs) or by Solar Energetic Particles (SEPs) (e.g.
\cite{Michel98}). However, such a production cannot account for the abundances of
extinct radionuclides derived from the observed isotopic anomalies mentioned above,
at least if the spallation processes have operated in the early solar system at a
level commensurable with their present efficiency. The only hope for a local
production to be viable is thus to call for an enhanced spallation production which
could be associated with the increased SEP production of the young Sun, especially
in its T-Tauri phase. The various studies conducted along these lines reach the
conclusion that it appears difficult to account for the production of the relevant
short-lived radionuclides in proportions compatible with the observations (e.g.
\cite{Podosek97} for references; also \cite{Sahijpal98}). 

If indeed the short-lived radionuclides that have been present live in the early
solar system are not of local origin, the message they carry on the chronology of
the nucleosynthetic events  responsible for a ``late pollution" of the solar system
can obviously not be extracted from the chemical evolution models needed when one
deals with long-lived chronometers. Instead, a scenario relying on a limited number
of events has to be constructed. Such a ``granular" model has been tailored by Dave
\cite{Schramm78}, and made more specific with the so-called ``Bing Bang" model
\cite{Reeves78,Reeves79}, which envisions the contamination and formation of the
solar  system in an OB association
during its approximate $10^7$ y lifetime. A chronology based on these granular
chemical evolution models raises a series of important and difficult questions
related in particular to the type of nucleosynthetic event(s) responsible for the
contamination, the corresponding radionuclide yields, as well as the efficiency of
the pollution. Dave has actively contributed at different levels to the scrutiny of
these problems (e.g.
\cite{Schramm78} for some references). He has in particular been very much concerned
with the possible role supernovae have played in the production of
short-lived radionuclides, as well as in their injection into the forming solar
system. Under Dave's supervision, Margolis \cite{Margolis79} has developed
hydrodynamical simulations of the contamination of a proto-solar type cloud by the
impacting (isotopically anomalous) gas or ``shrapnel-like" grains of a supernova
shell. At this occasion, grains have been shown to be more efficient contaminating
agents than the gas as a result of their higher probability of penetration in the
solar nebula, as well as of the reduced danger of having the isotopic anomalies
washed out beyond recognition in the bulk nebular material before the start of the
solar system solidification sequence (of course, this does not exclude grain
vaporisation {\it during} the solidification, which seems to be requested by
the analysis of the isotopic anomalies attributed to the radionuclide in situ decay).
The priviledged role possibly played by grains was most welcome at a time Dave and
his collaborators had studied in some detail the sequence of consensation of grains
around exploding stars \cite{Lattimer78}. At about the same time, the possible role
in the supernova contamination efficiency of ``fast moving knots" just discovered in
the supernova remnant Cas A was stressed \cite{Arnould78}, and may be nicely
complementary to the contaminating importance of grains. There is indeed mounting
evidence that fast moving knots are priviledged locations of grain formation in
supernova ejecta \cite{Lagage96}. Certainly, the details of the contamination are
still far from being settled. In any case, it seems highly plausible that the
short-lived radionuclides have been distributed  heterogeneously in the forming
solar system (see also \cite{Podosek97}). This complicated situation may affect quite
negatively their chronological predictive virtues. 

The possible role of supernovae in the pollution of the early solar system with
short-lived radionuclides (along with other stable nuclides) has continued to be
studied actively. Other types of contaminators have also been proposed, like novae
or Asymptotic Giant Branch stars. Here, and in the spirit of the Bing Bang model,
we briefly recall below that  massive stars of the Wolf-Rayet type
might also have been responsible for the production of a quite large suite of
short-lived radionuclides, and of their injection into the forming solar system.
We devote also a brief discussion to the short-lived radionuclides \chem{146}{Sm}
and \chem{205}{Pb} which have been explored by Dave, and which have been the
subject of some recent work. 

\subsubsection{WR stars: short-lived radionuclide contaminators of the early solar
system ?}

Wolf-Rayet (WR) stars are fascinating objects that have a dramatic impact on their
surroundings through their huge winds. They have been, and still are, the focus of
much observational and theoretical efforts. Their main properties have been reviewed
by \cite{Arnould97}, and are not repeated here.

The production of short-lived radionuclides of cosmochemical
interest by a large variety of WR stars has been calculated in the framework of 
detailed evolution models, and with the use of extended nuclear reaction networks
\cite{APM97}. As the modelling of these stars is immensely simpler than the one of all
the other short-lived radionuclide producers proposed up to now, the predicted
WR yields are thus likely to reach a level of reliability that cannot be
obtained in the other cases. In short, the main results obtained by \cite{APM97} are
as follows:
 
\begin{figure}[tb]
\centerline
{\bf
{\vbox{\psfig{figure=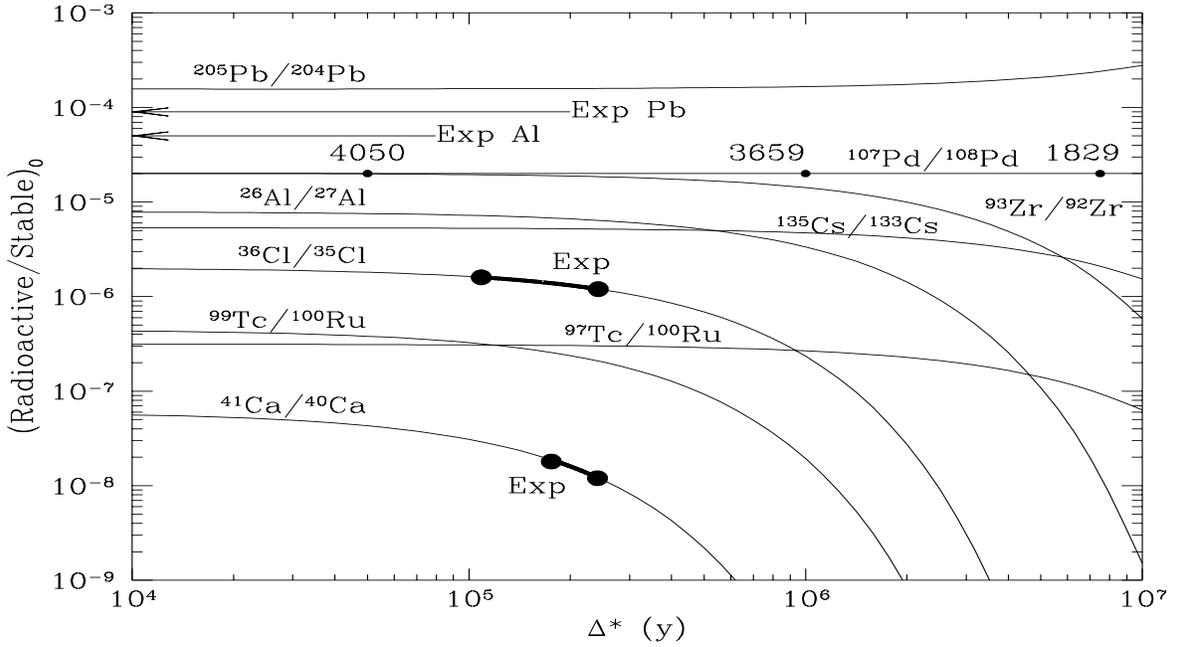,angle=270,height=9.0cm,width=\textwidth}}}
}
\caption[]{\small
Abundance ratios $\mbox{(R/S)}_0$ of various radionuclides R relative
to stable neighbours S versus $\Delta^\ast$ (see main text) for a
60 \ms$\,$ model star with $Z = 0.02$. All the displayed ratios are
normalized to
$(\chim{107}{Pd}/\chim{108}{Pd})_0 = 2\times10^{-5}$ (e.g. \protect\cite{W85})
through the application of a common dilution factor $d(\Delta^\ast)$. The
values of this
factor are indicated on the Pd horizontal line for 3 values of
$\Delta^\ast$. Other available experimental data (labelled Exp) are displayed.
They are adopted from \protect\cite{MacPherson95} for Al, \protect\cite{SUG94} for
Ca (see also \protect\cite{Sahijpal98}), \protect\cite{MGS97} for Cl, and
\protect\cite{Huey72} for Pb [see also
\protect\cite{Chen87}, who propose the somewhat larger value
(\chem{205}{Pb}/\chem{204}{Pb})$_0 \approx 3 \times 10^{-4}$ (not shown)] (see
\protect\cite{APM97} for more details)}
\end{figure}

\begin{list}{(\arabic{obj})}{\usecounter{obj}\setlength{\leftmargin}{0pt}
              \setlength{\labelwidth}{0pt}\setlength{\itemindent}{5pt}
              \setlength{\listparindent}{\parindent}
              \setlength{\parsep}{\parskip}
              }
\vskip-2truemm
\item The neutrons released by \reac{22}{Ne}{\alpha}{n}{25}{Mg} during the
He-burning phase of the considered stars are responsible for a s-type
process leading to the production of a variety of $A > 30$
radionuclides. In the absence of any chemical fractionation between the relevant 
elements, it is demonstrated that
\chem{36}{Cl}, \chem{41}{Ca} and \chem{107}{Pd} can be produced by this s-process in
a variety of WR stars of the WC subtype with different initial masses and compositions
at a {\em relative} level compatible with the meteoritic observations. For a
$60$ \ms $\,$ star with solar metallicity, Fig.~1 shows that this
agreement can be obtained for a time $\Delta^\ast \approx 2\times10^5$ y, where
$\Delta^\ast$ designates the time elapsed between the last
astrophysical event(s) able to affect the composition of the solar nebula and
the solidification of some of its material (e.g. \cite{W85}). More
details concerning other model stars are given by \cite{APM97};

\item To the above list of radionuclides, one certainly has to add \chem{26}{Al},
which is of crucial importance not only in cosmochemistry, but
also for $\gamma$-ray astronomy (Sect.~4). A detailed discussion of its
production by the MgAl chain of hydrogen burning has been conducted recently by
\cite{APM97} and \cite{MAPP97}. Let us simply note here that the canonical
value $(\chim{26}{Al}/\chim{27}{Al})_0 = 5\times10^{-5}$ \cite{MacPherson95}, while
not reached in the 60 \ms $\,$ star displayed in Fig.~1, can be obtained from the
winds of $M \geq 60$ \ms $\,$ stars with $Z > Z_\odot$ under the same type of
assumptions as the ones adopted to construct Fig.~1. Let us also note that the WR
models can account for the correlation  between \chem{26}{Al} and \chem{41}{Ca}
observed in some meteorites  \cite{Sahijpal98};
 
\item In  contrast, too little \chem{60}{Fe} is synthesized;

\item An amount of \chem{205}{Pb} that exceeds largely the experimental upper
limit set by \cite{Huey72}, but which is quite compatible with the value reported by
\cite{Chen87}, is obtained not only for the model star displayed in Fig.~1, but also
for the other cases considered by
\cite{APM97}. This high production appears in fact to be a distinctive prediction
concerning WR stars, and a renewed search for its in situ decay in meteorites would
be most valuable (Sect.~3.1.3);

\item More or less large amounts of \chem{93}{Zr}, \chem{97}{Tc},
\chem{99}{Tc} and \chem{135}{Cs} can also be produced in several cases, but
these predictions cannot be tested at this time  due to the lack of reliable
observations.
\end{list}

It has to be remarked that the above conclusions are derived without taking into
account the possible contribution from the material ejected by the eventual
supernova explosion of the considered WR stars. This supernova
might add its share of radionuclides that are not produced
abundantly enough prior to the explosion. This concerns in particular
\chem{53}{Mn}, \chem{60}{Fe} or \chem{146}{Sm}. One has also to acknowledge
that the above conclusions sweep completely under the rug the possible role of
binarity in the WR yields. Its impact on the predicted \chem{26}{Al}
production and the additional level of uncertainty it generates have been
explored by \cite{LBF95}.

 From the results reported above, one can try estimating if indeed there is
any chance for the contamination of the protosolar nebula with isotopically
anomalous WR wind material at an {\em absolute} level compatible with the
observations. In the framework of Fig.~1, this translates into the possibility
of obtaining reasonable dilution factors $d(\Delta^\ast)$. A qualitative
discussion of this  highly complex question based on a quite simplistic
scenario is presented by \cite{APM97}. In brief, it is concluded that
astrophysically plausible situations may be found in which one or several WR
stars with masses and metallicities in a broad range of values could indeed
account for some now extinct radionuclides that have been injected live into
the forming solar system (either in the form of gas or grains). Of course, a
more definitive conclusion would have to await the results of a more detailed
model that takes into account the high complexity of the WR circumstellar
shells, and of their interaction with their surroundings, demonstrated by
observation and suggested by numerical simulations. Concomitantly, the
possible role of WR stars, either isolated or in OB associations, as triggers
of the formation of some stars, and especially of low-mass stars, should be
scrutinized.           

\subsubsection{\chem{\rm 146}{\rm Sm}: a short-lived p-process radionuclide}

There is now strong observational evidence for the existence in the early solar
system of the two p-process radionuclides 
$^{92}{\rm Nb^g}\, (t_{1/2} = 3.6\, 10^7$ y) and 
$^{146}{\rm Sm}\, (t_{1/2} = 1.03\, 10^8$ y) (\cite{Harper96}, and references
therein). The case of $^{92}{\rm Nb^g}$ has been discussed by \cite{Rayet95}, who
conclude that various uncertainties in the level of production of this radionuclide
make rather unreliable at this time the development of a \chem{92}{Nb}-based
p-process chronometry.

As far as \chem{146}{Sm} is concerned, the study of its potential as a p-process
chronometer has been pioneered by Dave in collaboration with Audouze
\cite{AS72}. This work has triggered a series of meteoritic,
nuclear physics and astrophysics investigations, which have helped clarifying many
aspects of the question. In particular, the early uncertainties on the amount
of \chem{146}{Sm} that has been injected live into the solar system have been greatly
reduced through several studies of the \chem{142}{Nd} excess observed in certain
meteorites as the result of the in situ \chem{146}{Sm} $\alpha$-decay. More
specifically, it is concluded to-day that
(\chem{146}{Sm}/\chem{144}{Sm})$_0 = 0.008 \pm 0.001$, \chem{144}{Sm} being the
stable Sm p-isotope. One can attempt building up a p-process chronometry on this
value if the corresponding isotopic production ratio can be estimated reliably
enough at the p-process site.

Much work has been devoted to the modelling of the p-process in massive stars, and
especially in Type II supernovae (e.g. \cite{Arnould98}). In spite of this, the 
production ratio $P \equiv$ \chem{146}{Sm}/\chem{144}{Sm} remains quite uncertain,
being estimated by \cite{Somorjai98} to lie in the $0.7 < P < 2$ range in (spherically
symmetric) Type II supernova models. This unfortunate situation relates in
part to astrophysical problems, and in part to nuclear physics uncertainties,
especially in the \reac{148}{Gd}{\gamma}{\alpha}{144}{Sm} to
\reac{148}{Gd}{\gamma}{n}{147}{Gd} branching ratio (e.g. \cite{Rayet92}), even if the
prediction of this ratio has recently gained increased reliability. This
improvement comes from the direct measurement of the
\reac{144}{Sm}{\alpha}{\gamma}{148}{Gd} cross section down to energies very close to
those of direct astrophysical interest
\cite{Somorjai98}, complemented with a better nuclear reaction model
\cite{ArnouldHir98}. The resulting astrophysical rate is predicted to be 5 to 10
times lower than previous estimates in the temperature range of relevance for the
production of the Sm p-isotopes.  By application of the detailed balance theorem, the
rate of the reverse \reac{148}{Gd}{\gamma}{\alpha}{144}{Sm} of direct astrophysical
interest is reduced accordingly. This implies a lowering of the \chem{144}{Sm}
production, and favours concomitantly the \chem{146}{Sm} synthesis through the main
production channel
\chem{148}{Gd}($\gamma$,n)\chem{147}{Gd}($\gamma$,n)\chem{146}{Gd}($\beta^+$)\chem{146}{Sm}.
The net effect of the revised
\reac{148}{Gd}{\gamma}{\alpha}{144}{Sm} rate is thus an increase of the $P$ values.

Other nuclear problems add to the uncertainty in the evaluation of $P$. As noted
above, this concerns in particular the \reac{148}{Gd}{\gamma}{n}{147}{Gd} reaction,
for which no experimental information can be foreseen in a very near future in view
of the unstable nature of \chem{147}{Gd} ($t_{1/2} \approx 38$ h). An analysis of the
sensitivity of $P$ to this rate has been conducted by \cite{Rayet92} for SN1987A. 

In view of the difficulty of predicting $P$ reliably, one has to conclude that 
\chem{146}{Sm} cannot be viewed at this point as a reliable p-process chronometer. 

\subsubsection{\chem{\rm 205}{\rm Pb}: a short-lived s-process chronometer?}

Among the short-lived radionuclides of potential cosmochemical and astrophysical
interest, \chem{205}{Pb} ($t_{1/2} = 1.5 \times 10^7$ y) has the distinctive property
of being of pure s-process nature, at least if the \chem{204}{Tl} $\beta$-decay
competes successfully with its neutron capture in stellar plasmas. This remarkable
feature has not escaped Dave's attention, and he has pioneered with his collaborators
the study of the chronometric virtues of the \chem{205}{Pb} - \chem{205}{Tl} pair
\cite{Blake73,Blake75}. 

This early work has led its authors to express some
doubts about the possibility to rank \chem{205}{Pb} as a reliable s-process
clock. Apart from the fact that the level of the possible \chem{205}{Pb}
contamination of the early solar system was very poorly known (only an upper
limit of about $9 \times 10^{-5}$ being available for the
(\chem{205}{Pb}/\chem{204}{Pb})$_0$ ratio \cite{Huey72}), this pessimism related to
the realization that electron captures by the thermally populated 2.3 keV first
excited state of
\chem{205}{Pb} might reduce drastically the \chem{205}{Pb} effective lifetime in a
wide range of astrophysical conditions. Of course, the likelihood of a late
injection of \chem{205}{Pb} into the (proto-)solar nebula was reduced accordingly.

This conclusion has been demonstrated to be invalid, in certain s-process conditions
at least. The \chem{205}{Pb} destruction into \chem{205}{Tl} by electron captures may
indeed be efficiently hindered by the reverse transformation, which is made possible
as a result of the \chem{205}{Tl} bound-state $\beta$-decay. The nuclear aspects of
this question have been analyzed in considerable detail by \cite{Yokoi85}, who have
shown on grounds of schematic astrophysical models that the possible level of
\chem{205}{Pb} s-process production may be large enough to justify a renewed interest
for the \chem{205}{Pb} -- \chem{205}{Tl} pair. 

The work of \cite{Yokoi85} has indeed triggered further observational,
experimental and theoretical efforts. In particular, a new measurement of the
(\chem{205}{Pb}/\chem{204}{Pb})$_0$ ratio has been attempted, leading to a value of
about $3 \times 10^{-4}$ \cite{Chen87}. On the other hand, some experiments are
currently deviced in order to obtain a direct measurement of the \chem{205}{Tl} --
\chem{205}{Pb} mass difference with high precision \cite{Vanhorenbeeck98}. This
quantity, which is still somewhat uncertain, affects quite drastically the
predicted
\chem{205}{Tl} bound-state $\beta$-decay. Finally, more reliable estimates of the
\chem{205}{Pb} yields have been obtained through detailed s-process calculations
performed with the help of realistic model stars. This concerns in particular
Wolf-Rayet stars (see Sect.~3.1.1). Some estimates of the yields from thermally
pulsing Asymptotic Giant Branch (AGB) stars have also been made
\cite{Wasserburg94}, with special emphasis on the ability of \chem{205}{Pb} to
survive in neutron-free locations in between thermal pulses \cite{Mowlavi98}. The
latter work concludes that the chances for a significant \chem{205}{Pb} yield from
AGB stars are likely to increase with the stellar mass for a given metallicity, or
to increase with decreasing metallicity for a given stellar mass. However, the
modelling of the s-process in AGB stars still raises many questions which remain
to be answered before putting the \chem{205}{Pb} yields from these stars on a safe
footing.  

In short, one may conclude from the above considerations that much cosmochemical,
nuclear and astrophysics work remains to be done for giving a chance to
\chem{205}{Pb} to gain the status of a reliable short-lived s-process chronometer.

\subsection{Extinct short-lived radionuclides in the solar system}

The year 1987 has marked the tenth anniversary of the first discovery and isolation
of presolar grains in meteorites. This has been the start of a remarkable series of
dedicated laboratory work that has by now led to the identification and analysis of
a long suite of such grains, interpreted as specks of stardust having survived the
formation of the solar system. These presolar materials are refractories of various
types (diamond, SiC, graphite, corrundum, silicon nitride), some of them containing
even tiny subgrains, in particular, Ti-, Zr- and Mo-carbide or TiC subgrains in
graphite or SiC grains, respectively (see \cite{Bernatowicz97} for many
contributions on presolar grains).
All the analyzed elements contained in these
grains exhibit much larger anomalies than those found  in the  material that
condensed in the solar system itself. This is interpreted as the largely undiluted
nucleosynthetic  signature of specific stellar sources.

This rule applies in particular to the \chem{26}{Mg} excesses attributed to
the in situ decay of \chem{26}{Al} observed in presolar silicon carbide, graphite and
oxide grains, as demonstrated by e.g. Fig.~14 of \cite{MacPherson95} (see also the
reviews on specific grain types in \cite{Bernatowicz97}).
The initial \chem{26}{Al}/\chem{27}{Al} ratio inferred to have been present in the
analyzed grains vary from about $10^{-5}$ to values as high as about 0.5, to be
compared to the canonical solar system value (\chem{26}{Al}/\chem{27}{Al})$_0
\approx 5 \times 10^{-5}$ (Sect.~3.1.1). The highest reported ratios obviously put
particularly drastic constraints on the \chem{26}{Al} production models, especially
when the Al data are complemented with correlated isotopic anomalies in other
elements, and in particular in C, N, and O. As discussed in some detail by
\cite{Arnould97},  WR stars could well explain even the highest reported
\chem{26}{Al}/\chem{27}{Al} ratios, but might have some problem accounting for the
isotopic composition of, in particular, nitrogen. In addition, one has to
acknowledge that there is no clear indication at this point that the types of
grains loaded with large \chem{26}{Al} amounts can indeed condense from the WR
winds. 
 
Another example is provided by an extraordinary neon component, referred to
as Ne-E(L), which is carried by presolar graphite grains, and is made of almost pure
\chem{22}{Ne} (e.g. \cite{Amari95}). This remarkable feature is generally
interpreted in terms of the in situ decay of the ultra-short radionuclide
\chem{22}{Na} ($t_{1/2}
\approx 2.6$ y). In view of its short lifetime, the production of \chem{22}{Na} in the
thermonuclear framework requires the consideration of explosive situations. The first
explicit connection of this sort has been made by \cite{Arnould74} through detailed
explosive H burning calculations. They substantiated the later view
\cite{Clayton76} that Ne-E is hosted by nova grains.\footnote{The two forms of Ne-E 
identified to-day, Ne-E(L)
and Ne-E(H), were not known at that time. It is generally considered by now that
the less extreme \chem{22}{Ne} enrichments exhibited by Ne-E(H) do not require a
\chem{22}{Na}-decay origin}  Over the years, many
calculations have been carried out along these lines (e.g. \cite{Jose98} for a recent
study). Supernovae could also be
\chem{22}{Na} producers through explosive C burning, as demonstrated by the early
calculations of \cite{Arnett78}, and confirmed by more recent studies (e.g.
\cite{Woosley95}).

Some graphite grains and a couple of special SiC grains (referred to as SiC-X)  also
carry \chem{41}{K} excesses of up to two times solar that are attributed to the
in situ decay of
\chem{41}{Ca} ($t_{1/2} = 10^5$ y). From these observations, the intial \chem{41}{Ca}
abundances are inferred to lie in the $10^{-3} \lsimeq \chim{41}{Ca}/\chim{40}{Ca}
\lsimeq 10^{-2}$ range
\cite{Amari97}. The \chem{41}{Ca} production may be due to a s-process-type of neutron
captures associated with He burning in AGB \cite{Wasserburg95} or in WR
\cite{Arnould97} stars. However, the (uncertain)
\chem{41}{Ca} load of the AGB winds is predicted to be too low to
account for the observations. The situation is slightly more favorable in the case of
the WR stars, even if the highest observed ratios remain out of reach. Supernovae can
also eject some
\chem{41}{Ca} whose abundance relative to \chem{40}{Ca} can be of the order of
$10^{-2}$ in a variety of O- and C-rich layers
\cite{Woosley95}. A suite of isotopic anomalies accompanies the
\chem{41}{K} excess, and in particular an inferred \chem{26}{Al}/\chem{27}{Al} ratio
ranging typically between 0.01 and 0.1. Accounting for these correlated anomalies
may be a source of embarassment for the supernova scenario. Their proponents call in
particular for some large scale mixing of various and ad hoc amounts of different
supernova layers. 
 
Finally, evidence for the presence of \chem{44}{Ti} ($t_{1/2} \approx 60$ y) in some
graphite and SiC-X grains is provided by \chem{44}{Ca} excesses that translate into
\chem{44}{Ca}/\chem{40}{Ca} ratios of up to about 140 times solar \cite{Amari97}.
This radionuclide is predicted to be produced in the so-called $\alpha$-rich
freeze-out developing in the layers of a Type II supernova located just outside the
forming neutron star. The corresponding yields are thus extremely sensitive to the
still uncertain details of the physics of the explosion. It might also be produced in
some Type Ia supernova models where He detonation plays an important role (e.g.
\cite{Woosley95}, and references therein). The isotopic anomalies that correlate with
the \chem{44}{Ca} excess are generally considered to demonstrate a Type II supernova
origin of the carrier grains \cite{Nittler96}. Again, a large mixing of ad hoc amounts
of different supernova layers has to be postulated. The possibility  of getting just
the right mixing of course remains to be demonstrated.

\section{Gamma-ray lines from cosmic radioactivities}

As reviewed in the previous sections, short- or ultra-short-lived radionuclides
have left their signatures in solar system solids or meteoritic presolar
grains in the form of excesses of their daughter products. They may also be
identifiable in the present interstellar medium if their decay leads to a
substantial feeding of a nuclear excited state of the daughter products. In such a
situation, their electromagnetic de-excitation  produces $\gamma$-ray lines
with specific energies, usually in the MeV domain.

Gamma-ray astrophysics complements in a very important way the study of
extinct radionuclides in meteorites. In particular, it provides    information on the
present-day production of these nuclides; it also allows in some instances a
direct identification of their nucleosynthetic sources, while the
cosmochemical inferences are  necessarily indirect. In spite of this
complementarity, \ga-ray astrophysics is one of the rare astrophysical disciplines
that escaped Dave's interest.
Clayton and his collaborators pioneered
the study of the detectability of \ga-ray lines from the decay of radioactive nuclei
synthesized in supernova explosions, about  30 years ago.
This research was in fact the natural follow-up of the predictions
\cite{Bodansky68} that \chem{56}{Fe} is produced in a supernova as the
radioactive \chem{56}{Ni}, its \chem{56}{Co} decay product ($t_{1/2} \approx 77$
days) being in its turn responsible for the powering of the optical supernova light
curves.
This early work has been followed by the prediction that additional radionuclides
could be detectable as $\gamma$-ray line emitters (e.g. \cite{Clayton82}). With time,
it has also been realized that supernovae are  not the sole objects
of relevance in this field, and that various radionuclides could also be ejected from
novae, or even mass-losing AGB or WR stars.

On the observational side, $\gamma$-ray line astronomy has received considerable
momentum in the 80ies with the detection of \chem{26}{Al} in the Milky Way and
of \chem{56}{Co} in the  supernova SN1987A. In the  early 90ies, the
Compton Gamma-Ray Observatory (CGRO, with the OSSE and COMPTEL instruments in
particular) has contributed further to promote $\gamma$-ray line astronomy to a mature
astrophysical discipline (see  \cite{Diehl98,Prantzos99} for recent reviews).
  
\begin{table*}
{\small
\begin {center}
\caption{                COSMIC  RADIOACTIVITIES AND \ga-RAY LINES}
\begin{tabular}{ccccc}
\hline \hline
DECAY \ CHAIN &        LIFETIME$^*$          & LINE \ ENERGIES       &
 SITE  &  MAIN \\
              & (y) &  (MeV) & [Detected] & PROCESS \\
\hline \noalign {\medskip}
\nia \ra \co \ra \fe & 0.31  & {\underline {0.847}} (1) \
                   {\underline {1.238}} (0.685)  & SN  & NSE \\
  & & 2.598 (0.17) \ 1.771 (0.45)   & [SN1987A]    &    \\
                                        &       &    & [SN1991T]    &    \\
                                        &       &    &     &    \\
\ci \ra \fr &   1.1   & {\underline {0.122}} (0.86) \
                        {\underline {0.136}} (0.11)  & SN  & \aa-NSE  \\
            &       &    & [SN1987A]     &    \\
                                        &       &    &      &    \\
\na \ra \ne &   3.8   & 1.275 (1)                   & Novae & Ex H \\
                                        &       &    &      &    \\
\ti \ra \sca \ra \ca & 90 & {\underline {1.156}}(1)
& SN & \aa-NSE \\
& & 0.068 (1) \ 0.078 (0.98) & [CasA, GRO J0852-4642] &  \\
             &       &    &  &      \\
\alb \ra \mg &   $1.1 \times 10^6$   & {\underline {1.809}} (1)
   & WR, AGB          &  St H  \\
          &       &    &  Novae &   Ex H        \\
          &       &    &  SNII  &  St+Ex Ne   \\
          &       &    & [Galaxy]           &  \\
          &       &    &               &       \\
\fh \ra \ch \ra \nh & $2.2 \times 10^6$  &
1.322 (1) \ 1.173(1)  & SN   & n-NSE \\
\hline \hline
\end{tabular}
\end{center}
}
{\footnotesize *: For double decay chains the longest lifetime is given; \\
 $Underlined$: lines already detected; \\
 Numbers in $parentheses$: branching ratios;
In $brackets$: sites of line detection; \\
SN: supernova; SNII: Type II supernova \\
 St : Hydrostatic burning; Ex: Explosive burning \\
NSE: Nuclear statistical equilibrium;
 \aa: \aa-rich freeze-out;  n-: normal freeze-out
   }
\end{table*}

\subsection {Basics of \ga-ray line astronomy}

The most important radioactivities for \ga-ray line astronomy
are presented in Table~1, along with the corresponding lifetimes,
energies, branching ratios, main processes of production, and sites where \ga-ray
lines have been, or are expected to be, detected. It is seen that many of
the nuclides of cosmochemical interest discussed in the previous
sections (\chem{22}{Na}, \chem{26}{Al}, \chem{44}{Ti} and \chem{60}{Fe}) are also
$\gamma$-ray line radioactivities.

The production  of a radionuclide at a high enough level and its
decay to an excited state of its daughter nucleus are necessary but not
sufficient conditions for it to be an interesting candidate for \ga-ray astronomy.
Other factors indeed
play a key role. This concerns in particular the decay
lifetimes, which enter the problem through the fact that the
  production of the nuclei of interest  takes place in
environments initially opaque to \ga-rays; these photons are thus degraded in
energy as they interact with the  surrounding material. The
\ga-ray lines have in fact a significant probability to be detectable only if the
matter densities, and thus the opacities, become low enough on
timescales shorter than the radioactive decay lifetimes. These conditions can be met
in AGB or WR stars, which eject through extensive steady winds a substantial
fraction of their relatively low-density outer material that can be enriched with
certain \ga-ray line candidates. However, most radionuclides of interest are produced
in explosive events of the nova or supernova types. In the latter case
the synthesis of the nuclides of relevance takes place in highly opaque deep
layers, so that especially drastic constraints are put on the \ga-ray line
observability. The situation is, in fact, quite different when dealing with the
low-mass star explosions (Type Ia supernovae,  SNIa), or
with  explosions of massive stars (Type II supernovae, SNII). The SNIa ejecta
reach low opacites much more quickly than the SNII ones in view of their lower
masses ($\sim$1
\ms) and larger ejection velocities (in excess of $1.5 \times 10^4$ km/s). More
specifically, the ejecta become transparent to
\ga-ray photons  typically  after a few weeks in SNIa and only
after about one year in the SNII case. Even if SNIa provide better detection
conditions, they forbid in particular the observation of the \ga-ray line associated
with the decay of the  important nuclide \chem{56}{Ni}, whose half-life is
only 6 days.
 
The intensity of the escaping \ga-ray lines gives important
information on the yields of the decaying nuclides and on the
physical conditions (temperature, density, neutron excess, etc.) in their
production zones. In addition, the shape of the \ga-ray lines reflects the
velocity distribution of the ejecta, and can thus help probing its structure (e.g.
\cite{Burrows91}). Up to now, only the \coa \ lines from SN1987A and the \alb \ line
from the inner Galaxy have been resolved.

When the lifetime of a  radionuclide is not much larger than the time interval between
two nucleosynthetic events in the Galaxy, those events appear as \ga-ray
point sources. In the opposite case, a diffuse galactic emission is expected
from the integrated emission of many sources.  Characteristic timescales between two
explosions are about 1 to 2 weeks for novae, about 40 y for SNII+SNIb, and about
300 y for SNIa, these values being derived from the galactic frequencies of the
corresponding events \cite{DellaVale94,Tammann94}. A comparison between these
timescales and the decay lifetimes of Table~1 indicates that \alb \ and
\fec \  must be responsible for a diffuse emission, the spatial distribution of which
reflects the galactic distribution of the nucleosynthetic sources (except if high
velocity ejecta can travel undecelerated for long times; see Sec.~4.4).  The other
radioactivities of Table 1 should be seen as point sources in the Galaxy, except
perhaps $^{22}$Na if it is dominantly produced by ONeMg-rich novae. Indeed, about 40
of these objects might explode in the Galaxy over the $^{22}$Na lifetime.

\subsection {\chem{56}{Co} and \chem{57}{Co} lines from supernovae}

The observations in SN1987A of \ga-ray lines associated with the \chem{56}{Co} and
\chem{57}{Co} decays have confirmed in a spectacular way the predictions concerning
the synthesis of radioactive nuclei in supernovae. 
In particular, they have demonstrated that \fea, the most strongly bound stable
nucleus in nature is produced in the form of unstable \nia.
The main points of relevance to
\ga-ray-line astronomy are as follows:

\noindent (1) The detection of the 0.847 and 1.238 MeV \chem{56}{Fe} de-excitation
lines \cite{Matz88} about 6 months earlier than expected suggests that
the SN1987A ejecta has suffered large scale mixing and/or fragmentation during the
explosion or shortly after, bringing heavy nuclei from the inner layers into the
outer ones. This implies that the explosion does not
preserve the onion-skin structure predicted by 1-D pre-supernova model stars. 
Prompted by this observation, hydrodynamical 2-D and 3-D simulations showed that
mixing may indeed take place, due mostly to Rayleigh-Taylor type instabilities
(e.g. \cite{Fryxel91,Hashisu90}. This was a major contribution of
\ga-ray line astronomy  to the understanding of
supernova explosions; 

\noindent (2) The observed profiles of the 0.847 and 1.238 MeV lines are red-shifted
by 500-800 km s$^{-1}$ \cite{Tueller91}. This is at variance with the
expectations based on the hypothesis of an optically thick source. In addition, their
width is larger than predicted from theory, probably  indicating that some fraction
of \coa \ has penetrated  deeply into the high-velocity H-rich envelope. Despite some
preliminary models \cite{Burrows95,Grant93}, a convincing explanation of these data
does not exist yet;

\noindent (3) The features discovered at 122 and 136 keV \cite{Kurfess92} are
associated to the decay of \chem{57}{Co}. From the line intensities, the mass of
\chem{57}{Co} is estimated to be about $2.7 \times 10^{-3}$ \ms, which implies 
a \nib/\nia \ production ratio of $1.40 \pm 0.35$ times the solar ratio 
(\feb/\fea)$_{\odot}$ of the stable daughter nuclei. This suggests that most of the
\chem{57}{Co} has been produced in an ``\aa-rich'' freeze-out characteristic of a
relatively low-density environment
\cite{Clayton92}. The \chem{57}{Co} amount derived in such a way is compatible with
the one needed to account for the latest measured UVBRIJHK light curves of SN1987A
\cite{Fransson98}.

It has to be noted that the \chem{56}{Co} and \chem{57}{Co} line observations
reported above have been made possible because of the proximity of SN1987A. In
fact, these lines would have remained undetected with the available instruments if
the same event had occured in the Andromeda galaxy, and they would have been
just marginally detectable by an INTEGRAL-type instrument (to be launched in 2001).
This is due to the long timescale required for the slowly expanding massive
ejecta of SN1987A (of the SNII type) to become transparent to \ga-rays.
As discussed in Sect.~4.1, this attenuation
would have been much less drastic in a SNIa case. An additional factor in favour of
SNIa is that these explosions are predicted to produce about 0.5 to 1 \ms \ of
\chem{56}{Ni}, which is  typically ten times the average SNII
yield. In such conditions, the \coa \ lines from SNIa would be detectable up to the
Virgo cluster of galaxies (located at about 13-20 Mpc) by instruments with a
sensitivity of about $10^{-5}$ \cs, which is close to the sensitivity limit of CGRO.

Evidence for the  847 and 1238 keV \chem{56}{Co} lines has been
obtained by COMPTEL 66 and 176 days after the explosion of the
bright SNIa SN1991T in the spiral galaxy NGC4527, which is located at an estimated
distance of 17 Mpc at the periphery of the Virgo  cluster.  The obtained line flux 
\cite{Morris95} corresponds to more than 1.3 \ms \ of \nia \ if the distance 
exceeds 13 Mpc, implying that almost the entire exploding white dwarf responsible
for the supernova has been incinerated into
\nia. Sub-Chandrasekhar mass models for SNIa (with a detonation at the base of  the
accreted He layer inducing a further  detonation inside the white dwarf), or
delayed detonation models (where the flame front propagates subsonically at large
distances from the centre of the white dwarf before turning into a detonation) may
explain an early detection of the \ga-ray lines. However, they may have problems
accounting for the reported large \nia \ amounts. Before drawing firmer
conclusions, further detections of extragalactic SNIa are required, particularly in
order to clarify if indeed SN1991T can be considered as typical.

SN1991T illustrates the kind of diagnosis of SNIa models that can be achieved
through the analysis of their \coa \ lines. A detailed exploration
of the potential of this method has been performed recently
\cite{Gomez98,Hofflich98}. However, a
statistical analysis shows that the prospect of detecting SNIa with INTEGRAL
is rather dim, since its sensitivity to broad lines, such as those expected from the
high velocity SNIa ejecta, is not
much better than the CGRO one (for recent estimates, see \cite{Timmes98}).

\subsection{$^{44}$Ti lines from supernovae} 

As the \chem{44}{Ti} lifetime ($\sim$60 y) is
comparable to the characteristic timescale between two supernova explosions in the
Milky Way, the resulting \ga-ray line emission should appear as point sources. On
the other hand, this lifetime is sufficiently long for making \tii \ an excellent
probe of galactic supernova explosions in the past few centuries. In fact, its
\ga-ray lines might reveal supernova remnants (SNR) which have remained undetected up to
now at other wavelengths. This has been demonstrated by the recent detection of a
1.16 MeV emission from the previously unknown and presumably nearby SNR
GRO J0852-4642 \cite{Iyudin98}.

Cas A is one of the youngest (about 340 y old) and closest (about 3 kpc) known
supernova remnants. Optical and X-ray measurements suggest that its progenitor was a
20 \ms \ WR star that exploded as an underluminous supernova of the SNIb type. This
could explain why the explosion has remained undetected, despite its proximity and
high declination.  The 1.16 MeV \tii \ line has been detected in this remnant by
COMPTEL with a flux of ($4.2 \pm 0.9) \times 10^{-5}$ \cs \cite{Iyudin97}. This
translates into a \chem{44}{Ti} yield of about $1.5 \times 10^{-4}$ \ms, which is
not too far from the theoretical predictions. However, this detection has
brought forward an unexpected puzzle. Since \tii \ and \nia \ are synthesized in the
same stellar zones, the inferred amount of \tii \ has to be accompanied by about 0.05
\ms \ of \nia. If powered by the \nia \ decay, the supernova
(with or without a hydrogen envelope) would have a peak magnitude M$_V < -4$ at 3
kpc \cite{The95}. This implies that Cas A should have been a rather bright supernova
for a few weeks, making it difficult to understand why it went unreported. 
Among the various proposed solutions to the puzzle, the hiding of Cas A by a dusty
shell made of wind-ejected pre-supernova material appears as a quite natural
explanation, especially if the progenitor was a WR star 
\cite{Hartmann97}. Another interesting possibiity is that the deacy rate of \tii
\ was reduced during the early evolution of the CasA remnant (\tii \ decays by 
orbital electron capture and few electrons exist in an ionised environment). In that
case, the initial \tii \ mass may have been lower than implied by its currently 
observed activity \cite{Mochizuki99}.

Note that \tii \ is the third radioactivity that will eventually be discovered in
SN1987A. Along with \cob \ it is produced in the hottest and deepest
layers expelled by a SNII, so that the \tii \ and \cob \ yields are very sensitive
to the poorly understood physics of the explosion, and in particular to the position  
of the ``mass-cut'' (i.e. the
line dividing the supernova ejecta from the matter  accreted onto the compact
residue). Nucleosynthesis calculations for SN1987A predict the ejection of 
about $1.5 \times 10^{-4}$ \ms \ of \tii \  
(e.g. \cite{Thielemann96}). Comparable amounts are  suggested from
the fitting of the late  SN1987A  light curve if it  is indeed powered by \tii \
\cite{Fransson98}. 
Such amounts of \tii \ would produce on Earth a flux of  about 
$3 \times 10^{-6}$ \cs \ in the 68, 78 and 1157 keV lines
for many decades after the explosion. If \tii \ is ejected at low speed
(around $10^3$ km/s, as suggested by spherically symmetric models for SN1987A), the
kinetic broadening of the lines has to be small, and the estimated fluxes would be
close to the sensitivity limit of INTEGRAL for narrow lines. The detection of
\tii \ from SN1987A will certainly be a prime target
for that satellite, as it will provide a better probe than ever of the deepest
layers ejected by SN1987A.

\subsection {\alb \ and \fec \ (?) in the galactic plane}

The detection by HEAO-3 of the 1.8 MeV line associated with the decay of \alb \ 
\cite{Mahoney82} has marked the birth of the astrophysics of
\ga-ray lines. This observation provides an additional demonstration that
nucleosynthesis is currently active in the Galaxy,  and offers an interesting
opportunity to identify one of the sites of that activity (see \cite{Prantzos96a} for
a review of the \alb \ \ga-ray line astronomy and synthesis models).
 
\begin{figure}[tb]
\centerline
{\bf
{\vbox{\psfig{figure=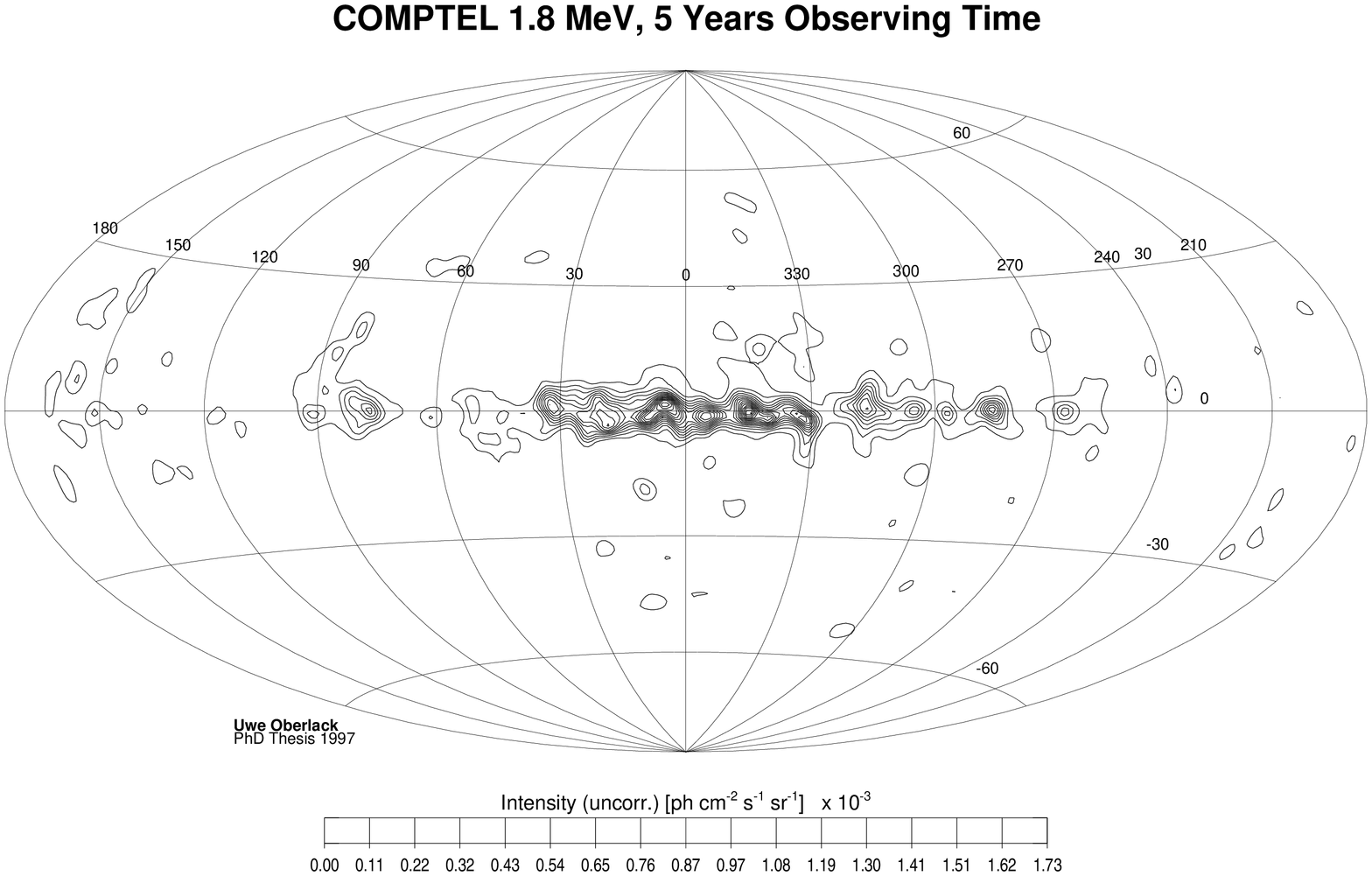,angle=270,height=9.0cm,width=\textwidth}}}
}
\caption[]{\small
Map of the Galaxy in the light of the 1.8 MeV line of $^{26}$Al (COMPTEL team).  }
\end{figure}

The COMPTEL \alb \ data reviewed by \cite{Diehl99} clearly exhibit a diffuse and
irregular 1.8 MeV flux along the galactic plane (Fig.2). 
These features provide an especially
good tracer of the current galactic \chem{26}{Al} production sites. In particular,
they exclude (i) a unique point source in the galactic centre, (ii) a strong
contribution from the old stellar population of the galactic bulge, and (iii) any
class of sources involving a large number of sites with low individual yields (novae,
low mass AGB stars), since a smooth flux distribution is expected in that case. In
contrast, they favour massive stars as the production sites for (most of) the derived
amount (about 2.5 \ms) of galactic \alb \ (e.g.
\cite{Prantzos93}). This scenario is made plausible in particular by the observed 1.8
MeV ``hotspots" with tangents to the spiral arms
\cite{Diehl95}\footnote{Note, however, that two of the COMPTEL 1.8 MeV hotspots
located  at the approximate galactic longitudes of $80^o$ and $90^o$ are
certainly not related to spiral features (Cygnus superbubble, Vela region). In
addition, the 1.8 MeV Vela hotspot is not associated with the nearby  Vela
supernova remnant alone, as originally  thought
\cite{Diehl99}}. An additional support comes from the
high level of similarity between a galactic map of the ionisation power from massive
stars and the 1.8 MeV map of the galactic \chem{26}{Al}. From these similar maps and
the choice of a standard initial mass function, \cite{Knodlseder99} is able to
reproduce the current galactic supernova rate and massive star population, and
suggests that most of the \alb \ is produced by high metallicity WR stars in
the inner Galaxy.

An intriguing recent observation concerns the spectral
width of the 1.8 MeV line from the inner Galaxy, which is estimated from measurements
by the GRIS spectrometer \cite{Naya96} to be $\Delta E = 5.4 \pm 1.4$ keV. This is
larger than the value of about 1 keV expected from the galactic
rotation. Even if \alb \ is initially ejected at high velocities, it is difficult to
understand how it could go undecelarated during most of its 1 My lifetime. Among the
possibilities explored by \cite{Chen97}, the condensation of \alb \ in
high-speed dust grains seems promising. However, the GRIS measurement needs to be
confirmed in the first place, since it is incompatible with the HEAO C line width
limit of 3 keV.

\def\ms{M$_{\odot}$}
\def\zs{Z$_{\odot}$}
\def\co{$^{56}$Co}
\def\nia{$^{56}$Ni}
\def\ti{$^{44}$Ti}
\def\fe{$^{60}$Fe}
\def\ga{$\gamma$}
\def\aa{$\alpha$}
\def\lr{$\rightarrow$}
\def\De{$\Delta E=\pm$}
\def\pcs{cm$^{-2}$ s$^{-1}$}
 
\begin{table*}
\begin{center}
{\small
\caption{                SOME PROSPECT FOR STELLAR \ga-RAY LINES WITH {\it INTEGRAL}}
\begin{tabular}{ccccc}
\hline \hline
ISOTOPE    & LINE E(MeV) & TARGET    & OBSERVABLE &  INTEREST \\
\hline
\hline \noalign {\smallskip}
 {\bf \co} & 0.847  & Extragalactic & Intensity  & Constrain models  \\
           & 1.238  & SNIa    & Shape      &  of SNIa          \\
\noalign {\smallskip}
\hline \noalign       {\smallskip}
          &   &  SN1987A    & Flux & Nucleosynthesis    \\
          &   &             &      & Mass-cut     \\
          & 0.068   &             &                         &             \\
{\bf \ti} & 0.078  &  CasA       & Confirmation            & $v_{ej}^*$\lr CasA Age  
\\
          & 1.156  &             & + Shape                 &  + Models    \\
          &   &             &                         &             \\
          &   & Galactic SN &  Intensity           &  Models +   \\
          &   &             &   + Shape            & Nucleosynthesis \\
\noalign  {\smallskip}
\hline \noalign       {\smallskip}
          &        &  Galaxy       & Accurate map     & Nucleosynthesis sites \\
          &         &              & Line width       &  Ejecta propagation   \\
{\bf \alb}&1.809   &              &                  &                        \\
          &   & Galactic          & Line Shape       & Distances          \\
          &   & Hotspots      & Flux            & Yields          \\
          &   &  (e.g. Vela )     & Extent           & $v_{ej}^*$          \\
\noalign    {\smallskip}
\hline \noalign        {\smallskip}
  {\bf \fe}& 1.173 & Galaxy   & Flux  & Sources    \\
           & 1.322  &               &  (map ?)        &                        \\
\noalign     {\smallskip}
\hline \noalign          {\smallskip}
  {\bf \na}& 1.275 & Novae    & Flux, Shape           & Models     \\
\noalign {\smallskip}
\hline \hline
\end{tabular}
}
\end{center}
{\footnotesize *: $v_{ej}$: speed of SN ejecta
   }
\end{table*}

An exciting perspective for INTEGRAL is the possibility to detect a diffuse
$\gamma$-ray emission at  1.2 and 1.3 MeV associated with the decay of $^{60}$Fe, and
give some hints about its distribution in the Galaxy. Indeed, the SNII models of
\cite{Woosley95} predict a \fec \ to \alb \ yield ratio of about 0.25-0.35 when
averaged over a reasonable stellar initial mass function. Taking into
account the respective \fec \ and \alb \ lifetimes (Table 1), the \fec \ line flux 
is expected to be about 0.15 times lower than the \alb \ one if indeed SNII are the
major galactic \alb \ producers. If INTEGRAL is unable to detect the \fec \ lines, 
SNII might have to be discarded as major \alb \ sources. At this point, it may be of
interest to emphasize that the \fec \ SNII yields depend upon
the details of the explosion model, and are thus still uncertain. On the other hand,
WR stars are also
\alb \ and \fec \ producers during their non-explosive and explosive (SNIb/c) phases,
respectively. This might complicate the interpretation of a possible \fec \ detection
by INTEGRAL.
 
\section {Summary and perspectives}

The decay of a variety of nuclides  recorded in live or fossil
form in the solar system brings a rich variety of information about a vast diversity
of highly interesting astrophysical questions. Dave Schramm has written important
pages of many chapters of this ``astrophysics of cosmic radioactivities" with his
incomparable dynamism and visionary perception of astrophysics. He has without doubt
triggered much work in the field and shown many new ways. For sure, he would be most
gratified to know that so much remains to be done and said, and that  each day brings
its share of renewed excitement and laboratory discoveries.

Gamma-ray line astronomy has become rightly one of the important components on
the scene, complementing nicely the information on cosmic
radioactivities gained from cosmochemical studies. The already observed \ga-ray lines
from \coa, \cob, \tii \ and \alb \ have allowed to probe various interesting aspects
of supernova explosions, and to better define the level of success, and also of
failure, of simple supernova models. The \alb \ emission line has also helped
locating sites of active large scale nucleosynthesis in the Galaxy. The future of
this young astronomy is bright, with the much awaited launch of INTEGRAL, a
\ga-telescope with improved sensitivity, to be launched be ESA by the year 2001.
In Table 2 we present a brief synopsis of some
stellar nucleosynthesis problems that could be tackled with that instrument.

\end{document}